\documentclass{sig-alternate-05-2015}
\pdfoutput=1
\usepackage{todonotes}
\usepackage{amsmath}

\begin{document}

\setcopyright{acmcopyright}

\doi{10.475/123_4}

\isbn{123-4567-24-567/08/06}

\conferenceinfo{SIGIR PIR '16}{July 21, 2016, Pisa, Italy}

\acmPrice{\$15.00}

%
\conferenceinfo{SIGIR PIR}{'16 Pisa, Italy}

\title{Privacy Leakage through Innocent Content Sharing\\in Online Social Networks}
%
%
%
%
%

\numberofauthors{2} 
%
\author{
%
%
\alignauthor
Maria Han Veiga \\
       \affaddr{Inst. of Computational Science}\\
       \affaddr{University of Zurich, Switzerland}\\
       \email{hmaria@physik.uzh.ch}
\alignauthor
Carsten Eickhoff \\ 
       \affaddr{Dept. of Computer Science}\\
       \affaddr{ETH Zurich, Switzerland}\\
       \email{ecarsten@inf.ethz.ch}
}
\date{30 July 1999}

\maketitle
\begin{abstract}
The increased popularity and ubiquitous availability of online social networks and globalised Internet access have affected the way in which people share content. The information that users willingly disclose on these platforms can be used for various purposes, from building consumer models for advertising, to inferring personal, potentially invasive, information. 

In this work, we use Twitter, Instagram and Foursquare data to convey the idea that the content shared by users, especially when aggregated across platforms, can potentially disclose more information than was originally intended.

We perform two case studies: First, we perform user de-anonymization by mimicking the scenario of finding the identity of a user making anonymous posts within a group of users. Empirical evaluation on a sample of real-world social network profiles suggests that cross-platform aggregation introduces significant performance gains in user identification. 

In the second task, we show that it is possible to infer physical location visits of a user on the basis of shared Twitter and Instagram content. We present an informativeness scoring function which estimates the relevance and novelty of a shared piece of information with respect to an inference task. This measure is validated using an active learning framework which chooses the most informative content at each given point in time. Based on a large-scale data sample, we show that by doing this, we can attain an improved inference performance. In some cases this performance exceeds even the use of the user's full timeline.

\end{abstract}

%
%
%
%

%
%
    \printccsdesc



\section{Introduction}
\label{sec:introduction}
User privacy is a topic that has increasingly gained traction with the rise of online social networks (OSN). These platforms allow users to communicate, connect with peers and share content. Originally, OSNs mainly focused on these core aspects, but nowadays the term also includes platforms which are primarily user-centric, allowing members to broadcast personal thoughts and content. In 2010,~\cite{browsingkumar2010} find OSNs among the most frequently visited Web sites for a large population of users. Due to their prevalence and abundance in personal content, OSNs lend themselves to the study of human behavior at scale~\cite{socialnetworkslazer2009}.

Recent successful initial public offerings (IPO) and high market valuations underline the monetary value of OSNs. However, the relation between the number of registered users, their online activity, and these valuations is not entirely clear. It has been shown in several studies that user characteristics, such as personality traits~\cite{kosinski2013fb} or future route intentions~\cite{visitseickhoff2012}, can be reliably inferred from corresponding OSN profiles. Although the general value of personal data is widely accepted, there have not been many studies which assign a tangible value to OSN profiles. As a consequence, both for users as well as platform providers, the value of information remains a vague notion, at best. This situation is detrimental both to users who cannot be expected to make informed decisions about privacy controls, as long as they do not know the value and potential risk of disclosing a given information item, as well as platform and service providers who blindly buy and sell user data in bulk instead of saving resources by concentrating on select relevant portions of information.

In this paper, we aim to draw attention to accidental privacy leakage through content sharing in online social networks and make a first step toward describing a formal metric of task-specific informativeness of pieces of shared content.

Our empirical study relies on three popular OSN platforms: (1) Twitter, a microblogging platform whose main content comes in \emph{tweets}, posts limited to 140 characters which can contain text, media (video or images), links to external Web sites, references to other users and \emph{hashtags} (terms starting with the \# symbol, which are used to mark keywords or topics in a tweet). (2) Instagram, a photo sharing platform. Its main content are visual in nature along with optional textual descriptors. (3) Foursquare, a location service platform concentrating on the notion of \emph{check-ins}. Check-ins correspond to real-world venues that the user has visited. In addition to the venue name, more information such as location and venue categories are available. 

Our investigation is driven by the following research questions:
\begin{enumerate}
    \item How well can we uniquely identify a user based on matching a set of unseen posts to a user's online footprint and is there a benefit in modelling user identities across more than one OSN?
    \item For the same user, is the information posted in one OSN indicative of the information contained in another OSN?
    \item Can we quantify the amount of new information that a piece of shared content carries, with respect to a concrete inference task?
\end{enumerate}

In particular, to address RQ 1, we mimic the following scenario: let us have a collection of users' online footprints and a set of $q$ anonymous posts. We test whether we are able to correctly find the author of the anonymous posts based on seeing part of their online footprint.

As for RQ 2 and 3, we consider that a user may unintentionally expose personal information through seemingly innocuous shared content. For example, when a user shares a venue check-in, it is easy to infer which venue category was visited. However, a post which does not mention a place explicitly might still contain information about a potential behaviour or visit intention. Consider two tweets from the same user: ``\textit{Lol should start heading to the gym \#fitness}'' and ``\textit{What a great sunny day!}''. It is clear that the first tweet contains more information about the user's intention to visit a venue type than the second. To this end, we devise an informativeness metric for shared content. The metric explicitly models the item's relevancy towards a given inference task as well as its novelty in comparison to the previously seen timeline. Such a score can serve as an indication of the amount of novel information disclosure associated with an information item and can, in the future help both service providers as well as privacy advocates in making informed decisions.

This paper makes three novel contributions beyond the state of the art: 
\begin{enumerate}
    \item We formulate a scoring function to quantify the information value of shared content from the perspective of improving the performance of a concrete inference task.
    \item We present a practical way of exploiting the theoretical model by integration into an active learning scheme that results in enhanced user model learning rates.
    \item In both practical settings, we put particular emphasis on cross-platform models of user identity
\end{enumerate}

The remainder of this paper is structured as follows: Section~\ref{sec:related} gives an overview of related work dedicated to user modelling as well as privacy protection in Web scenarios. On the basis of a parallel corpus of OSN profiles belonging to the same natural person, Section~\ref{sec:case1} discusses unique user identification methods employing intra-platform as well as cross-platform information. Section~\ref{sec:case2} formally describes an informativeness score for shared OSN content and applies it to the task of predicting user traits manifested on one platform based on the user's activity on other OSNs. Finally, Section~\ref{sec:conclusion} concludes with a brief discussion of our main findings as well as an outlook on ongoing and future extensions to this work.

\section{Related Work}\label{sec:related}
There is a wealth of work dedicated to privacy protection in Web information systems such as search engines or social networks. A number of early studies investigate the common privacy concerns of information system users~\cite{smith1996information,stewart2002empirical,malhotra2004internet,hong2013internet}, finding that general concerns are abundant among Internet users but remain vague and imprecise. Many users are aware of the information collection and behavioral profiling activities undertaken by service providers as well as the wide range of data-driven inference efforts that have been presented by the academic and industrial communities~\cite{visitseickhoff2012,kosinski2013fb}. In spite of this knowledge, however, even technology-affine users cannot reliably quantify the exact risks entailed by careless information disclosure.

Privacy concerns become especially prevalent in mobile computing environments~\cite{keith2013information}. De Montjoye \textit{et al.}~\cite{de2013unique} show that as few as four hourly GPS samples are enough to uniquely identify 95\% of all individuals in a 500k-user phone log. We encounter an even greater potential for privacy hazards in settings that go beyond raw positional traces, joining them with topical information, \textit{e.g.}, in Web search queries~\cite{west2013here} or contextual advertising~\cite{agarwal2011location}.

To counter such de-anonymization and tracking efforts, various strategies have been proposed. Dwork's concept of differential privacy~\cite{dwork2011differential} considers adding $\epsilon$-noise to aggregate queries that prevents singling out individual contributions to the overall aggregate. Similarly, Carpineto and Romano~\cite{carpineto2013semantic} rely on the notion of $k$-anonymity, ensuring that no query should return less than $k$ individual records. In the domain of personal information, these approaches may not go far enough since certain, frequent, characteristics that would neither be detected under $k$-anonymity nor differential privacy could cause severe privacy hazards. 

This paper, in spirit, follows the reasoning of Howard \textit{et al.}~\cite{howard1966information} by measuring not just the amount of information contained in a given message, but also with respect to an inference task which can be economically relevant. In this way it attributes an economic dimension to messages, which can be an interesting measure both for industry players as well as for the message's original author. On the basis of a number of concrete classification tasks, this paper aims to close the gap between the rich body of work on empirical analysis of privacy hazards on the one hand and the large range of available privacy protection measures on the other. We argue that only by understanding the concrete implications of information disclosure (\textit{e.g.}, in the form of the value of a piece of information) can users be expected to make educated decisions about the appropriate protection measures they are willing to take.

\section{Dataset}
\label{sec:dataset}
As our research questions are concerned with the relationship between parallel user profiles of the same natural person across different OSNs, we rely on the methodology described in~\cite{veigacross2016} to assemble our dataset. We obtain a collection of 618 distinct users who cross-post content from corresponding profiles in multiple social networks, totalling 1.1 million tweets, 18000 Instagram posts and 99000 Foursquare check-ins.

\section{User De-anonymization}
\label{sec:case1}

Our first use case is concerned with, given a number of anonymous social media posts $q$ and a collection of users $\mathcal{U}$, finding the particular user $u \in \mathcal{U}$ that authored the posts. Inspired by general text matching strategies~\cite{Berger2001}, each user's known previous posts are described in the form of a unigram language model $M_u$ and the likelihood of said user having authored the anonymous text $q$ corresponds to $p(M_u|q)$. Using Bayes' law, one can write:

\begin{equation}
    p(u|q) = \frac{p(q|u)p(u)}{p(q)}
\end{equation}

And to select the most likely user:

\begin{equation}
    \arg \max_{u \in \mathcal{U}} p(q|u) 
\end{equation}

To simplify the expression further, we assume that $p(q)$ is constant for all users and treat $p(u)$ as uniform across all $u\in\mathcal{U}$. Thus, we find the most likely user by estimating $p(q|u)$, the probability of posts $q$ being generated by the language model derived from $u$'s available timeline.

These timelines are projected into a n-dimentional TF-IDF weighted vector space. To preserve the natural way in which users write, no further vocabulary pre-processing (such as lemmatization or exclusion of less common words) was applied. Based on this representation, we estimate $p(q|u)$ as the product across all terms $t$ in the vocabulary:

\begin{equation}
    p(q|u) \propto \prod_{t\in V}p(t|M_u)
\end{equation}

As described in Section~\ref{sec:dataset}, the dataset contains the online footprint of the same user on Twitter, Instagram and Foursquare. To mimic the described task, textual data from one OSN is used as the source of anonymous posts and the textual data from the remaining two OSNs is used to generate the user language model $p(M_u|q)$. 

To generate the training data, randomly sampled sections of varying length from the training source are used to generate pairs of the form $(M_u, user)$. For the test set, we remove any form of user mentions to mimic an anonymous post.


Our experiments investigate a number of combinations of (training, test) data sources:
\begin{enumerate}
\item (Twitter, Twitter)
\item (Twitter + Foursquare + Instagram, Twitter)
\item (Twitter, Foursquare)
\item (Twitter + Instagram, Foursquare)
\item (Twitter, Instagram)
\item (Twitter + Foursquare, Instagram)
\end{enumerate}

For the first two cases, we split the Twitter timeline into separate training and test portions. However, the more interesting conditions are 3-6, as the source of anonymous posts does not come from the data source used to generate the language model.

Additionally, we vary the amount of available profiling information by successively revealing larger parts of the training data. Furthermore, we also study the impact of changing the size of the set of anonymous posts $q$.

Multiple training sources are combined (Conditions 2, 4, and 6) in the following way: for a fixed amount of available profiling information, 20\% of it is made up from Instagram or Foursquare data (or 40\% if these are put in together) and the rest from Twitter, due to the relative abundance of Twitter data.

The performance of our classifier is given by accuracy of predicting the correct user who generated the anonymous posts. The results are averaged across 10 randomization runs.



\begin{figure}
  \centering
      \includegraphics[height=2in]{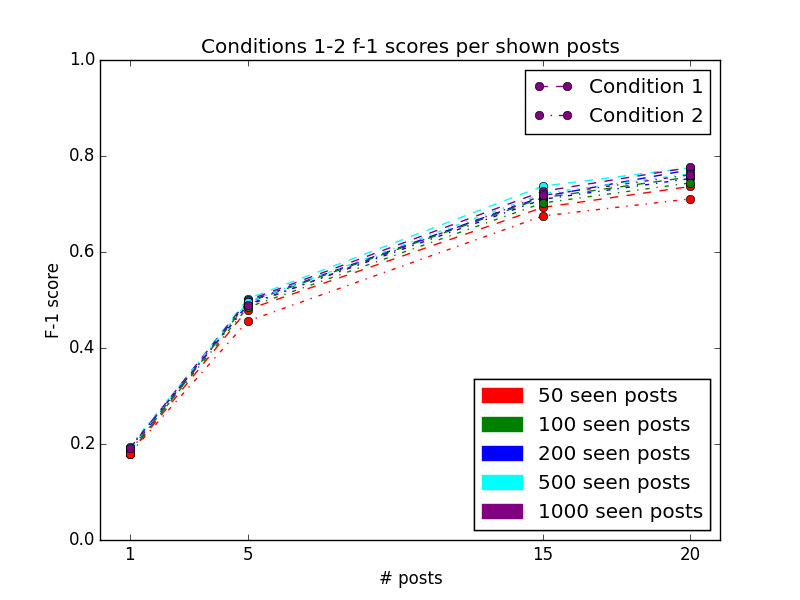}
  \caption{F-1 score curve of classifiers for conditions 1-2 varying sample size and training data size}\label{fig:f112} 
\end{figure}

\begin{figure}
  \centering
      \includegraphics[height=2in]{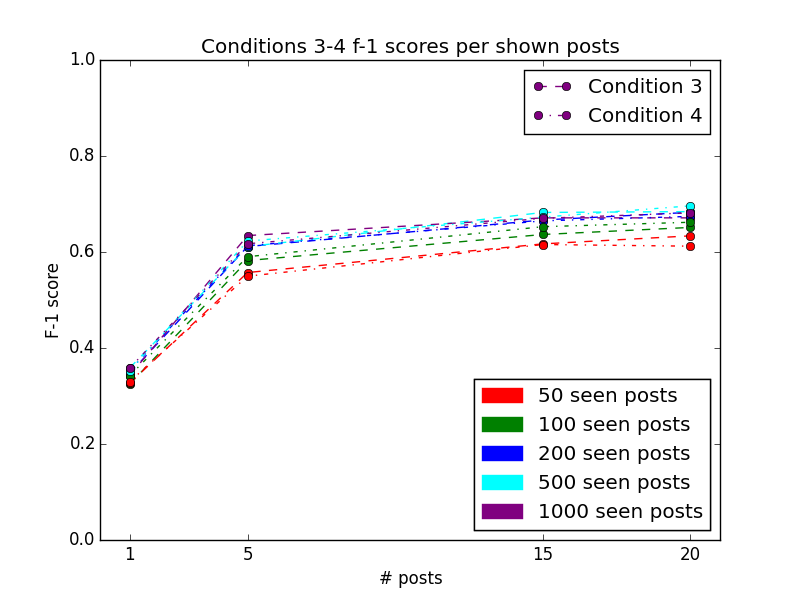}
  \caption{F-1 score curve of classifiers for conditions 3-4 varying sample size and training data size}\label{fig:f134} 
\end{figure}

\begin{figure}
  \centering
      \includegraphics[height=2in]{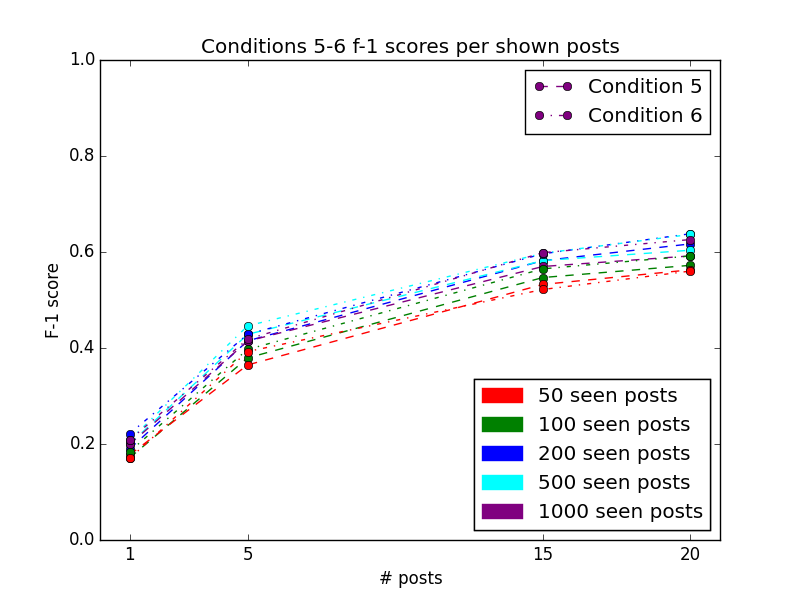}
  \caption{F-1 score curve of classifiers for conditions 5-6 varying sample size and training data size}\label{fig:f156} 
\end{figure}

In Figures~\ref{fig:f112}-\ref{fig:f156}, the micro averaged F-1 score curves for the classifiers built under different conditions are shown.

In Table~\ref{table:deanon}, the results for Conditions 1-2 can be found. We note that the usage of additional OSNs does not improve the de-anonymization performance. This is not too surprising as the source of the anonymous posts come from Twitter. The results for Conditions 3-4 can be found in Table~\ref{table:sourcefsq}. We remark that the classifier's performance does not improve with the addition of Instagram data as users can cross-post check-ins on Twitter and users can check-in into a venue multiple times.

The more interesting results can be found in Table~\ref{table:sourceinsta}, which presents the results for Conditions 5-6. We note that in this case, as training and test data come from different sources, there is some improvement in the de-anonymization performance when we include Instagram data as an extra training source.


\begin{table}[!htb]
    \caption{User de-anonymization for different sizes of training and test data, using Twitter as a source for anonymous posts}
      \centering
        \scalebox{0.9}{
        \begin{tabular}{| c |c | c | c | c | c |} 
 \cline{3-6}
 \multicolumn{2}{l}{} & \multicolumn{4}{|c|}{ \# anonymous posts}\\[2pt] 
 \hline
Train Source & Posts Seen  & 1 & 5 & 15 & 20\\
\hline
\multirow{5}{*}{Twitter}& 50 & \textbf{18.21} & \textbf{47.80 } & 69.90  & \textbf{74.22}  \\
& 100 & \textbf{19.19} & \textbf{48.85} & \textbf{71.41} & \textbf{76.96} \\
& 200 & 18.45 & \textbf{48.68} & 71.57 & 77.03\\
& 500 & \textbf{19.59} & 49.32 & \textbf{73.23} & \textbf{77.96} \\
& 1000 & \textbf{20.24} & \textbf{49.83} & \textbf{73.34 }& \textbf{78.23} \\
\hline
\multirow{5}{*}{\parbox{1.5cm}{Twitter \\Instagram Foursquare}}& 50 & 17.56 & 46.17 & \textbf{69.82 }& 73.24  \\
& 100 & 18.38 & 47.88 & 71.24 & 75.82  \\
& 200 & \textbf{18.78} & 48.35 & \textbf{72.23} & \textbf{77.69} \\
& 500 & 18.16 & \textbf{50.84} & 72.97 & 77.87 \\
& 1000 & 19.90 & 49.39 & 72.81 & 76.66 \\
[1ex] 
\hline
\end{tabular}}
\label{table:deanon}
\end{table}

\begin{table}[!htb]
    \caption{User de-anonymization for different sizes of training and test data, using Foursquare data as a source for anonymous posts}
      \centering
        \scalebox{0.9}{
        \begin{tabular}{| c |c | c | c | c | c |} 
 \cline{3-6}
 \multicolumn{2}{l}{} & \multicolumn{4}{|c|}{ \# anonymous posts}\\[2pt] 
 \hline
Train Source & Posts Seen  & 1 & 5 & 15 & 20\\
\hline
\multirow{5}{*}{Twitter}& 50 & 32.43  & 54.97  & 61.33  & 62.34  \\
& 100 & \textbf{33.17} & 58.35 & 63.24 &  63.10  \\
& 200 & 34.51 & \textbf{60.85} & \textbf{67.40} & 67.89  \\
& 500 & 35.81 & \textbf{62.41} & 67.50 & 68.94  \\
& 1000 & \textbf{36.91} & 61.58 & \textbf{67.88} & 67.06 \\
\hline
\multirow{5}{*}{\parbox{1.5cm}{Twitter Instagram}}& 50 & \textbf{32.68} & \textbf{55.72} & \textbf{61.87} & \textbf{62.80} \\
& 100 & 33.16  & \textbf{58.43}  & \textbf{64.14}  & \textbf{66.01}  \\
& 200 & \textbf{36.56} & 60.34  & 66.84 &\textbf{69.02} \\
& 500 & \textbf{37.58} & 62.24 & \textbf{69.17} & \textbf{69.04}  \\
& 1000 & 35.00 & \textbf{62.57} & 66.92  & \textbf{67.95} \\
[1ex] 
 \hline
\end{tabular}}
\label{table:sourcefsq}
\end{table}

\begin{table}[!htb]
    \caption{User de-anonymization for different sizes of training and test data, using Instagram data as a source for anonymous posts}
      \centering
        \scalebox{0.9}{
        \begin{tabular}{| c |c | c | c | c | c |} 
 \cline{3-6}
 \multicolumn{2}{l}{} & \multicolumn{4}{|c|}{ \# anonymous posts}\\[2pt] 
 \hline
Train Source & Posts Seen  & 1 & 5 & 15 & 20\\
\hline 
\multirow{5}{*}{Twitter}& 50 & 17.44 & 35.12 & 49.22 & 52.32 \\
& 100 & \textbf{19.07} &\textbf{40.17}  & \textbf{55.51} & \textbf{59.82} \\
& 200 & 20.77 & 41.71 & 58.06 & 60.07 \\
& 500 & 20.75 & 41.27 & 58.34 & 61.74 \\
& 1000 & 21.00 & 40.15 & 55.62 & 60.34 \\
\hline
\multirow{5}{*}{\parbox{1.5cm}{Twitter Foursquare}}& 50 & \bf{18.05} & \bf{35.91} & \bf{51.20} & \bf{53.17} \\
& 100 & 18.28  & 38.44 & 55.01 & 57.12 \\
& 200 & \bf{21.70} & \bf{44.81} & \bf{60.90} & \bf{64.50} \\
& 500 & \bf{22.55} & \bf{44.02} & \textbf{60.40} &\textbf{63.78} \\
& 1000 & \bf{21.87} & \bf{41.85} & \textbf{60.42} & \textbf{63.26} \\
[1ex] 
 \hline
\end{tabular}}
\label{table:sourceinsta}
\end{table}


With respect to RQ 1, we note that it is possible to match user profiles across OSNs based only on their textual data, achieving a maximum accuracy of \textbf{77.96}\% when using only 500 posts (the equivalent of 20\% of the average length of the Twitter timelines at our disposition). Furthermore, the usage of multiple OSNs as training data source seems to improve the classifiers' performance when the source of anonymous posts and training data are distinct, suggesting there is a consistency in user language and vocabulary across the chosen OSNs. We also observe that, in general, the more anonymous posts are available, the better the performance of the designed classifier becomes.

\section{Information Valuation}
\label{sec:case2}
Let us again start from an OSN user base $\mathcal{U}$, in which each user $u$ is defined by the set of his associated timelines $\{M_u^k\}_{k=1}^K$. Further, let $S$ be an OSN such that we can define the set of all posts made by $u$ in $S$ as his timeline, $M_u^S$. We treat the timeline as a long consecutive piece of text in which each post constitutes a sentence. We use information from the timeline to estimate the probability of a user manifesting a certain property $A$. This probability is denoted by $p( A | M_u)$, where $A$ denotes ``\emph{$u$ shows Property $A$}'' and $M_u$ is the user's timeline. Due to our definition of timelines, the same method can be used for full timelines or subsets of posts. Regardless of the chosen scope, we now project the timeline into an $n$-dimensional TF-IDF weighted vector space that allows us to train a classifier $\mathcal{C}_A$, estimating the final $p( A | M_u)$. 


\subsection{Measuring informativeness}
\label{sec:info}
Our objective is to find a function which quantifies the information carried in a post. On the one hand, we are interested in capturing the relevance of a post with respect to a certain inference task, on the other hand, in order to avoid redundancy or attributing a high score to already seen information, we are interested in capturing the novelty of some content with respect to what is already known. In a spirit similar to \cite{clarke2008}, we model the information content in two ways:
\begin{itemize}
\item Relevance $\rho$ of the post with respect to an inference task or a set of tasks;
\item Novelty $\nu$ of the post with respect to the user's previously posted content.
\end{itemize}

We form our informativeness score as a convex combination between these two quantities, thus introducing a mixture parameter $\lambda \in [0,1]$. Now, for each newly authored post $m$, we can define an informativeness function $\mathbb{I}: \mathbb{R}^n \times \mathbb{R}^{n} \times \mathcal{C} \to  \mathbb{R}^+$ as follows:

\begin{equation}
\begin{split}
\mathbb{I} (m, M_u, \mathcal{C}) &= \lambda \ \nu(m, M_u)\\
&+ (1-\lambda) \ \rho(m , \mathcal{C})
\end{split}
\end{equation}

\subsubsection{Relevance}

Measuring the relevance of shared content can be intuitively thought of as determining which piece of shared content contains features that are important for the classifier's decision. A popular choice of such a function describing feature importance is the \emph{Gini Importance} ($Ig$). For a feature $\theta$, the Gini Importance for a classifier $\mathcal{C}$ is defined as:

\begin{equation}
Ig_{\mathcal{C}} ( \theta ) = \sum_T \sum_\tau \Delta{i_\theta} (\tau, T).
\end{equation}

Where $\tau$ is a node, T a decision tree and $\Delta_i(\tau)$ the decrease in Gini Impurity. The Gini Importance indicates how often a particular feature $\theta$ was selected for a split, and how large its overall discriminative value was for a particular classification problem. We estimate the overall relevance $\rho$ of a post by summing up the importance scores across features contained in the post $\vec{m} \in \mathbb{R}^n$:
\[
\rho(\vec{m}, \mathcal{C}) \ = \sum_{i=0}^n Ig_c (m_i)
\]

\subsubsection{Novelty}
For a fixed $u \in \mathcal{U}$ and OSN $S$, let $\vec{m}_1$, $\vec{m}_2 \in \mathbb{R}^n$ be the vector representation of shared contents $m_1$, $m_2 \in M_u^S$, the user's timeline.

Informally, the function $novelty: \mathbb{R}^n \times \mathbb{R}^n  \to \mathbb{R}^+$ should have the following properties:
\begin{enumerate}
\item $\vec{m}_2$ should have low novelty if it is contained in $\vec{m}_1$.
\item $\vec{m}_2$ should have low novelty if it is similar to $\vec{m}_1$.
\item $\vec{m}_2$ should have high novelty if it is distinct from $\vec{m}_1$.
\end{enumerate}

Let $\vec{m_1}$ be denoted as $(m_{1,1}, ..., m_{1,n})$. The proposed function to measure novelty is the following: Let $\nu$: $\mathbb{R}^n \times \mathbb{R}^n  \to \mathbb{R}^+$, be a non-symmetric function defined by:

\begin{equation}
\nu (\vec{m_1}, \vec{m_2}) = \frac{\sum\limits_{i=1}^n \mbox{exp}(-\alpha (|m_{i,1}| + |m_{2,i}|-1))}{\sum\limits_{i=1}^n {1} [m_{2,i} \neq 0 ]}
\end{equation}

For each word, the novelty function decays with the number of times that the word appears. By regulating $\alpha$ we can control how many times we have to observe a word to not consider it novel anymore.

\subsection{Experimental Setup}
As a concrete example of the general data-driven label prediction problem introduced previously, we turn towards the task of predicting whether a person will visit a particular type of location (\textit{e.g.}, an Italian restaurant or a golf course) based on their social network timeline(s). These timelines are projected into a TF-IDF-weighted vector space. The vocabulary is curated by: removing all links and user mentions, stop words, words which occur less than 5 times and, when possible, word lemmatisation using WordNet. For the prediction task, we use the AdaBoost algorithm~\cite{breiman1998class} with decision trees as weak learners as our classifier since they generally work well without refined parameter tuning. The classifier's performance is evaluated under 10-fold cross-validation.

We begin by training one binary classifier per venue type that decides whether or not a user's timeline suggests they are likely to visit that type of location. For every test user $u$, we initiate the procedure by randomly sampling a single post from their timeline $S_k$ and create a truncated timeline. Then, at each iteration, we sample a constant number $d$ of posts from the timeline, add them to the truncated timeline and make a prediction using this iteratively updated input vector. The procedure is iterated until user $u$ has no more posts left (or until the truncated timeline reaches a fixed amount of posts). We obtain an ordered sequence of predictions: $(y_{0} , y_{1}, ... , y_{n_{end}})$, where $n_{end}$ represents the number of iterations. When $d=1$, \textit{i.e.}, we add one post at a time, we simulate the situation in which an existing user profile is updated over time as new content is being shared.

We aggregate results taking into account the varying timeline length across users in the following way: the maximum timeline length $n_{l_{max}}$ is calculated. Then, for each user whose timeline is shorter than $n_{l_{max}}$, the last prediction $y_{{n_{end}}}$ is repeated to generate a sequence of predictions of length $n_{l_{max}}$. As a performance baseline, we randomly sample posts to be added. In the active method, instead, we select the posts which are most informative according to our metric presented in the previous section.

\begin{figure}
  \centering
      \includegraphics[height=1in]{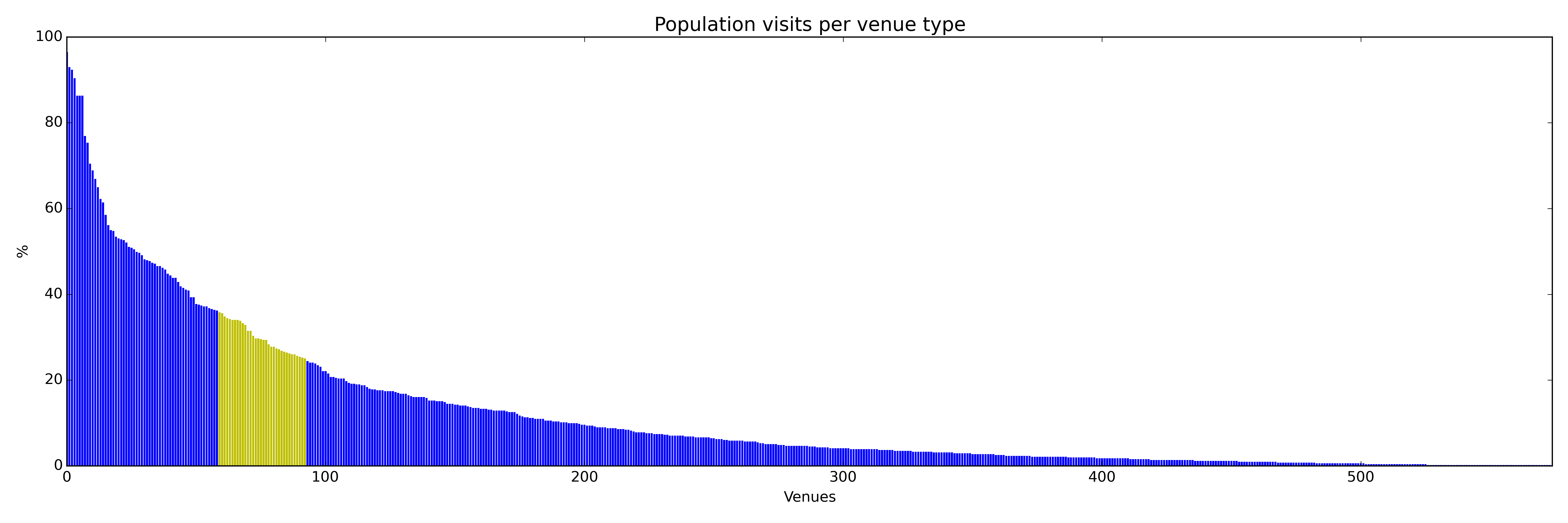}
  \caption{Percentage of users visiting different venue types}\label{fig:uservisits} 
\end{figure}

 Foursquare offers a hierarchical taxonomy of places that display very different relative popularity. Figure~\ref{fig:uservisits} plots the percentage of our user population that has visited each of the more than 500 categories. We can note that the distribution is heavily skewed. While virtually all users have, at some point, visited places that fall into broad categories such as [\textit{Arts \& Entertainment}] or [\textit{Food}], others are so specific that they remain almost empty (\textit{e.g.}, [\textit{South Tyrolean Restaurants}] or [\textit{Hunting Supplies}]). For the purpose of venue prediction, we are forced to make a subselection of categories that are neither so broad that the prediction task would become trivial nor so specific that the classifier would not find sufficiently many training examples. For this reason, we focus on those categories that were visited by 25\% to 35\% of the population, giving us a set of 37 venues (highlighted in yellow in the figure).

\subsection{Results}

\subsubsection{Cross OSN}
\begin{figure} \centering
 \begin{tabular}{@{}c@{}}
 \includegraphics[height=2.5in]{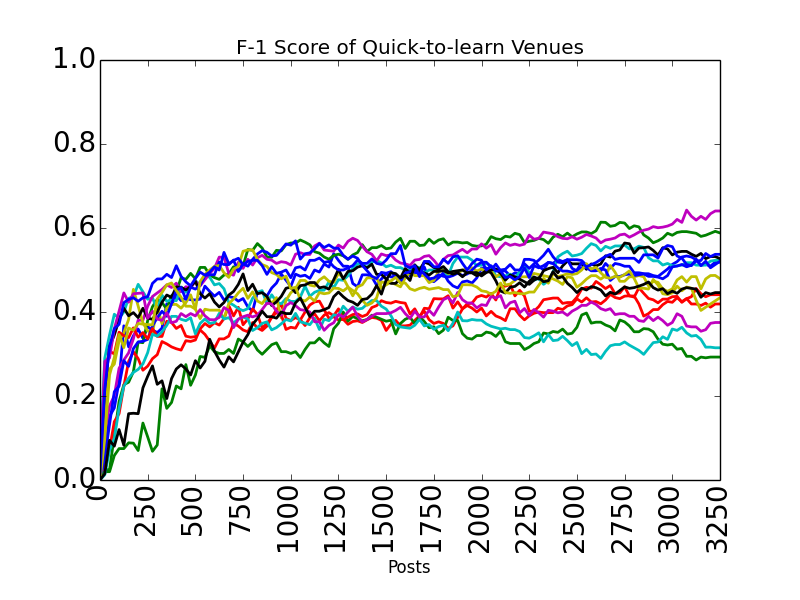} \label{fig:quick} \end{tabular}
  \begin{tabular}{@{}c@{}}
\includegraphics[height=2.5in]{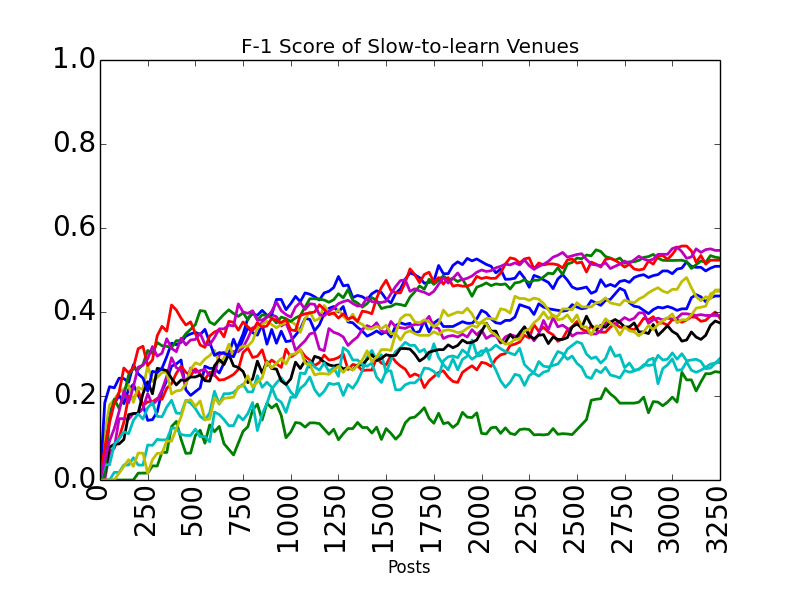} \label{fig:slow} \end{tabular}
 \begin{tabular}{@{}c@{}}\includegraphics[height=2.5in]{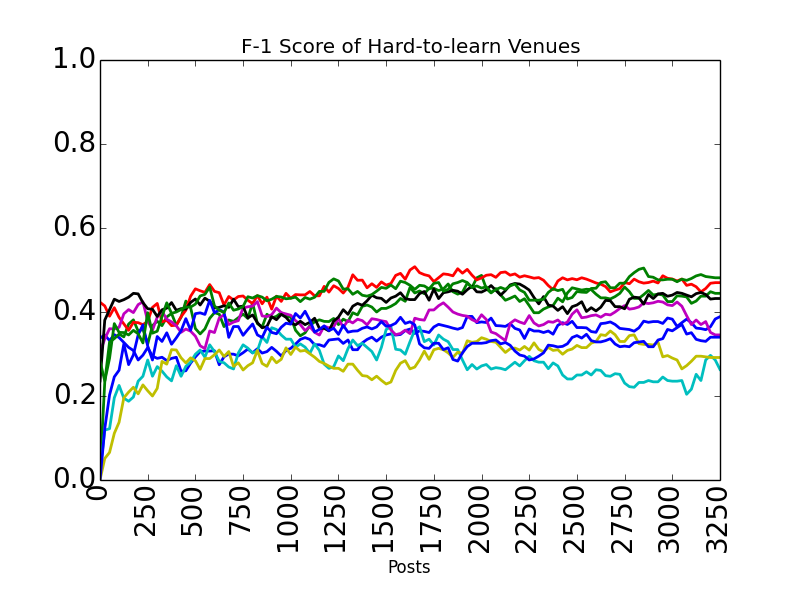}\label{fig:hard} \end{tabular} 
\caption{Venue-dependent classifier learning rates across training iterations.}\label{fig:learningvenues} \end{figure}

\begin{figure*}
\centering
\includegraphics[height=2.5in]{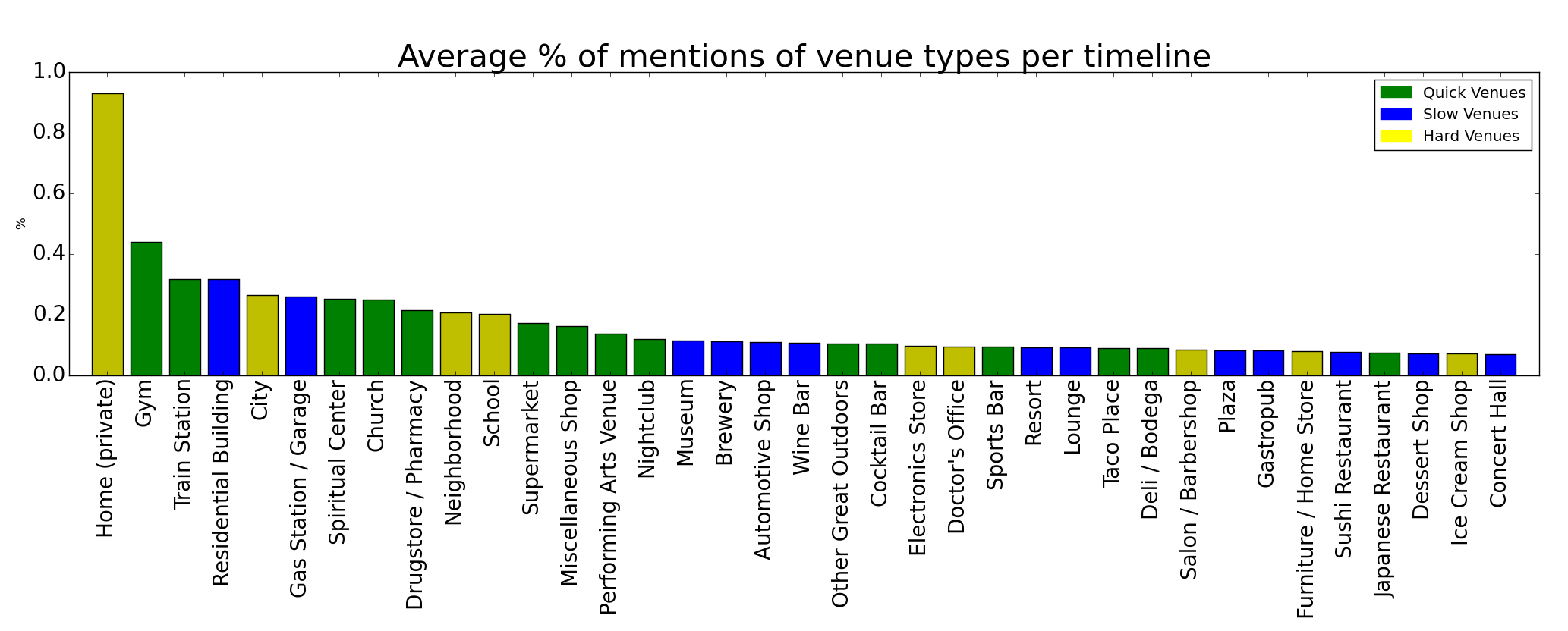}
\caption{Average percentage of venue type mentions per twitter timeline, when user has visited the venue type with category assignment of quick, slow or hard to learn for venue types.}\label{fig:hardeasy} 
\end{figure*}

Each user's timeline is initialized with a single post to which we iteratively add additional, randomly sampled, posts to form an updated user model. Figure~\ref{fig:learningvenues}, shows the classifier performance as a function of the number of posts available to the user model. From this overview, we note three recurring slope patterns. Some venue-specific classifiers quickly reach their optimal performance after as few as 750 posts have been observed (top figure), for others, significantly more iterations are required (center figure), and lastly, for some particular venues, the classification accuracy hardly benefits at all from using more posts (bottom figure). We refer to these three situations as quick-to-learn, slow-to-learn and hard-to-learn venues, respectively. Figure~\ref{fig:hardeasy} gives a complete overview of the relative frequency of mentions of the chosen venue categories and their affiliation to the three slope types. The general tendency seems to be that frequently mentioned venues tend to be quicker to learn than rare ones, while hard-to-learn venues appear to be randomly spread across the observation frequency range.

\subsubsection{Active resource selection}
After having confirmed the intuitive assumption that (within the limits of our three slope types) more data results in more accurate predictions, we now proceed to describing an active selection scenario in which we expand the user model by the most informative posts according to our metric rather than random ones. To this end, we fix the novelty parameter $\alpha$ at $0.5$, meaning that after a word appears 5 times, its novelty becomes negligible. Table~\ref{table:activelearning} highlights this method's performance at different settings of $\lambda$. 50 posts are actively selected for this experiment and we note that our selection scheme biased towards informativeness delivers significantly better $F_1$ performance\footnote{Average $F_1$ score for all classifiers is computed using the average of precision and recall across all classifiers.} than the random selection baseline at all parameter settings. The overall best performance was obtained at a setting of $\lambda = 0.1$, where the score mixture is dominated by feature importance while still taking into account novelty.


\begin{table}[h!]
\centering
\caption{Average classifier performance across all 37 venue types, for different $\lambda$.}
\scalebox{0.9}{\begin{tabular}{|c | c | c | c | c | c | c |} 
 \hline
$\lambda$ & $F_1$-score & Precision & Recall\\
\hline 
baseline & 15.36 & 52.78 & 8.99\\
0.0 &  42.10 &  44.21 &  40.18 \\
0.1 &  \textbf{44.19} &  \textbf{46.64} &  \textbf{41.97} \\
0.2 &  44.16 &  46.57 &  41.98 \\
0.3 &  43.63 &  46.00 &  41.48 \\
0.4 &  43.66 &  46.09 &  41.47 \\
0.5 &  43.23 &  45.50 &  41.18 \\
0.6 &  43.80 &  46.31 &  41.56 \\
0.7 &  43.55 &  46.85 &  40.69 \\
0.8 &  42.78 &  45.72 &  40.19 \\
0.9 &  42.21 &  45.99 &  39.00 \\
1.0 &  18.95 &  43.38 &  12.12 \\
[1ex] 
 \hline
\end{tabular}}
\label{table:activelearning}
\end{table}

\begin{figure}
\centering
 \begin{tabular}{@{}c@{}}\includegraphics[height=2.1in]{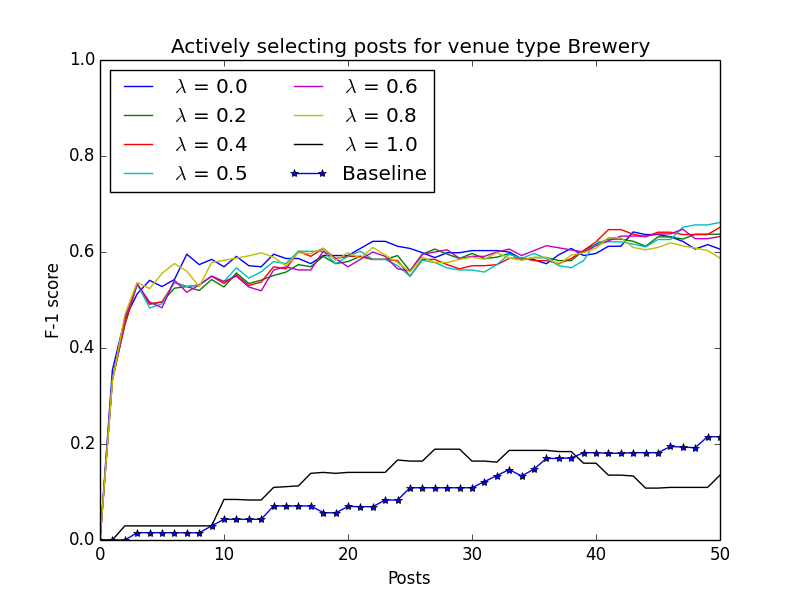} 
\end{tabular}
 \begin{tabular}{@{}c@{}} \includegraphics[height=2.1in]{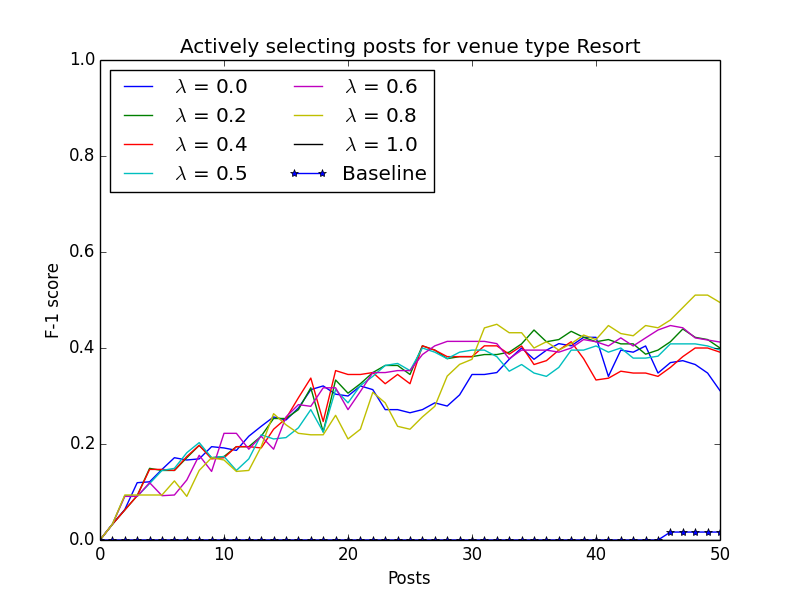} 
\end{tabular} 
\caption{Example of venue types which benefit from novelty component. Top figure: venue type \emph{Brewery}. Bottom figure: \emph{Resort}}\label{fig:difflambdas} 
\end{figure}

Let us return to our previously introduced categorization of learning curve slope types. Table~\ref{table:noveltymatters} shows the influence of $\lambda$ on the performance of the three slope categories. We observe clearly diverging tendencies between quick and slow-to-learn venues. While quick-to-learn venues benefit from low novelty contributions, their more slowly evolving counterparts benefit from novelty-biased informativeness scores. Examples can be seen in Figure~\ref{fig:difflambdas}. Again, hard-to-learn venue types do not show any noticeable response to different choices of $\lambda$, as long as the relevance component is not fully turned off.

\begin{table}[!htb]
    \caption{Average $F_1$-scores across slope types.}
      \centering
        \scalebox{0.9}{
        \begin{tabular}{|c | c | c | c |} 
 \hline
$\lambda$ & Quick & Slow & Hard\\
\hline 
0.0 &  47.81 & 38.51 & 36.70 \\
0.2 &  \textbf{50.16} & 40.40 & 38.43 \\
0.4 &  49.83 & 40.10 & 37.23 \\
0.6 &  49.59 & 40.18 & \textbf{38.48} \\
0.8 &  46.97 & \textbf{41.19} & 36.97 \\
1.0 &  17.87 &  9.70 & 30.48 \\
[1ex] 
 \hline
\end{tabular}}
\label{table:noveltymatters}
\end{table}

Furthermore, for some particular venues, the classifier attained a better performance when using only 50 actively selected posts than when using the full timeline of the users. Some examples can be found in Table~\ref{table:truncatedvsfull}.

Regarding RQ 2, we show again that there is consistency in terms of content shared across OSNs, in particular, we show that is is possible to predict venue type visits based on what is shared on Twitter and Instagram. Furthermore, with respect to RQ 3, we show that using our designed metric, we can find the posts which are most relevant to predict venue type visits. In particular, using the active learning framework with our information measure of content in posts as selection criteria, we observe overall quicker learning rates and in some cases, we can use a significantly reduced the number of posts to attain a classifier performance which is comparable to using the full timeline of the user.

\begin{table*}
\centering
\caption{Classifier performance using a truncated timeline versus a full timeline.}
\scalebox{0.9}{\begin{tabular}{|c | c | c | c | c | c |} 
 \hline
 & Sushi Rest. & Cocktail Bar  & Gastropub  & Brewery & Nightclub \\
\hline 
Full timeline & 52.86 & 31.48 & 38.53 & 50.88 & 43.40\\
Truncated + Random Selection & 11.49 & 4.97  & 5.36 & 21.51 & 20.11\\
Truncated + Active Selection & 61.72 & 55.84 & 54.79 & 66.15  & 54.23\\
[1ex]
 \hline
\end{tabular}}
\label{table:truncatedvsfull}
\end{table*}

\section{Conclusion}\label{sec:conclusion}
In this paper, we studied privacy hazards pertaining to cross-platform social network usage. Individually innocuous posts can lead to leakage of critical information when aggregated along or across a user's OSN profiles. We quantify this effect in two experiments: (1) uniquely identifying users in an anonymous pool and (2) predicting user properties manifested on one OSN platform based on content from other parallel profiles. 

In the user de-anonymization task, we note that it is possible to match user profiles across OSNs based only on their textual data, with as little as 10\% to 20\% of the user's full timeline.  Furthermore, the inclusion of multiple OSNs as training data sources has been shown to improve the classifiers' performance when the source of anonymous posts and training data are distinct. This suggests that there is a consistency in user language and vocabulary across OSNs.

In the information valuation task, we propose a general-purpose metric of textual informativeness in order to model the value of shared information items both for service providers (predictive power) as well as the user (potential privacy hazards). We show experimentally that the metric reflects the relative importance of posts with respect to the inference task being performed. When actively selecting a subset of posts per user, this method was always able to beat a random selection baseline. While choosing posts according to their relevance seems to lead to better performance in general, we noted that only for some venues there was a noticeable benefit in including a strong novelty component in the information scoring function. 

This work focused on showing the privacy hazards that arise from sharing content which seems uninformative or harmless. In the future, we are excited to extend this line of work by a dedicated investigation of information valuation scores on the user side (\textit{e.g.}, of an OSN) as it would greatly help people understand their own digital footprint and enable them to recognize moments of critical information disclosure. Furthermore, part of this work focused on proposing a metric for information valuation with respect to an inference task. We are interested to extend this line of work by an \textit{in-vivo} study of monetary efficiency of advertisers as a consequence of introducing an informativeness-aware resource selection scheme in their real-time bidding (RTB) pipelines.


\bibliographystyle{plain}
\bibliography{sigproc.bib}  
\end{document}